# Suppressing Rayleigh-Plateau Instability with a Magnetic Force Field for Deformable Interfaces Engineering

Arvind Arun Dev*[1,2], Thomas Hermans[3] and Bernard Doudin[1]

[1]Université de Strasbourg, CNRS, IPCMS UMR 7504, 23 Rue du Loess, 67034 Strasbourg, France

[2]School of Applied and Engineering Physics, Cornell University, Ithaca, NY 14853, USA.

[3] Université de Strasbourg, CNRS, UMR7140, 4 Rue Blaise Pascal, 67081 Strasbourg, France

The Rayleigh-Plateau instability (RPI) is a classical hydrodynamics phenomenon that prevents a jet of liquid to flow indefinitely within air or another liquid. Here, we show how adding a magnetic force field makes possible its suppression. Enclosing the jet in a ferrofluid held by magnetic forces allows flow focusing without sheath flow, which completely avoids dripping failure at small flow rates and provides conditional stability for a continuous fluid jet. Highly deformable liquid interfaces withstanding spatial and time varying flow conditions within a large parameter space can be realized.

RPI phenomena[1,2] sets severe limits on the stability of liquid-liquid and liquid-air interfaces, stating that their deformation leads to irreversible instability of the liquid flow. This manifests as an inevitable breaking of a continuous liquid jet into droplets[3] or dewetting of solid surface coating [4]. RPI is put to use in printing, pesticides spraying, fog harvesting and even for prey hunting by the Archer fish [5]. There is however a strong need to find strategies to overcome RPI, for example for drug delivery with lubricated flow [6], optimization of hydropower[7], and delicate bio material manipulation[8], where highly flexible interfaces are mandatory.

Soft and deformable liquid interfaces are nevertheless abundantly found in nature. Air-water interface that provides habitat for water striders[9], elastic salvinia surface in presence of water drop[10] and blood carrying elastic arteries[11] are a few examples. Biomimicking coupled fluids elasticity behavior have been studied in the form of soft lubrication[12], elastocapillarity [13] viscous peeling[14], and electro-elastocapillarity[15]. The physical understanding of a hydrodynamics-mediated deformable interface and the related phenomena is of paramount importance to mimic natural systems[16–18], make analogs of electronic circuits[19–21] and promise technological application [22–24] like sensors, flow devices and microfluidic circuits, taking advantage of a controlled soft wall confinement. Deformable interfaces also govern fundamental hydrodynamic issues; from the suppression of flow fingering instability [25] to triggering the laminar to turbulent transition in tubular flow [26].

We show how adding a magnetic contribution to the potential energy of a flowing liquid allows conditional stability against RPI, resulting in a new type of soft interface. It relies on our demonstration that an appropriate design of magnetic forces can stabilize a ferrofluid liquid envelope around an inner immiscible liquid[27]. We detail here the fundamentals of the response of a liquid-in-liquid flow interface under dynamic conditions, measure its deformation in response to pressure driven flow, and make explicit its stability criteria, revealing the remarkable large parameter space where flow stability is maintained.

A liquid-in-liquid cylindrical tube[27], called 'antitube'[28] is experimentally stabilized by a quadrupolar source magnetic field generated by four permanent magnets, of length $L$ (along $z$) of several cm, much larger than their mm range separation $W$, as sketched in Fig. 1 (inset Fig. 1a)). A liquid of small or negative magnetic susceptibility (hereby referred as 'non-magnetic') defines the antitube of diameter $D$, enclosed within an immiscible envelope made of ferrofluid with comparably much higher magnetic susceptibility (hereby referred as 'magnetic'). The permanent magnets generate a magnetic field (see supplementary information S1 for magnetic field calculation details) that pulls the ferrofluid away from the center and stabilizes the antitube in their center, where the magnetic field is negligible. The excess pressure inside the antitube is, $P = P_L + P_M = \sigma\kappa + \left(\frac{1}{2}\mu_0 M_I^2 + \int_0^{H_{Max}} \mu_0 M(H) dH\right)$, with the Laplace pressure $P_L$ resulting from the interfacial tension $\sigma$ at the liquid-liquid interface (see supplementary information S2 for interfacial tension measurements), with curvature under small slope assumption $\kappa = \frac{1}{R} - R''$, (R is the radius of the antitube and ' denotes the derivative with respect to the flow direction $z$). The pressure is augmented by the occurrence of Maxwell magnetic stress on the magnetic liquid $P_M$, sum of the magnetic traction at the magnetic-nonmagnetic interface $\frac{1}{2}\mu_0 M_I^2$, with $\mu_0$ the magnetic permeability, $M_I$ the magnetization at the interface that follows a cylindrical symmetry in first approximation, and the bulk magnetic pressure $\int_0^{H_{Max}} \mu_0 M(H)dH$ [29]. The latter results from the ferrofluid magnetization $M(H)$, up to $H_{Max} = 2M_r D/\pi W$ the approximated magnetic field generated by a quadrupolar source magnets of remnant magnetization $M_r$ [27]. The ferrofluid magnetization as a function of applied field follows a Langevin curve (see supplementary information S2), approximated by a linear function at low *H* and a constant saturated value at larger applied field. While specific values depend on the ferrofluid properties, the linear relation holds ($M = \chi H$) for small antitube diameter, as the applied field goes to zero when $D$ becomes negligible, and the saturation approximation ($M = M_s$) becomes relevant for large diameters, when *D* approaches *W*. Fig. 1a) illustrates the saturated and linear cases, where magnetic pressure increase with the antitube diameter, in contrast to the decreasing Laplace pressure (cf. dashed line). For the static no-flow condition, the balance of these two pressures results into a stable liquid-in-liquid structure, of antitube equilibrium diameter $D_{eq}$, where $P_L = P_M$. For $D > D_{eq}$, the pressure of magnetic origin dominates the destabilizing Laplace pressure, which is the core of the RPI stabilization phenomenon.

Insight into antitube size and shape was obtained by X-ray absorption contrast imaging[27,30] techniques in 2D using a homemade X-ray setup[27] and in 3D using a RX-Solutions Easy Tom 150/160 tomographic setup (see supplementary information S3 for experimental set up). Fig. 1b) shows the 3D tomography image of a deformed antitube under flow $Q$ = 75 µL/min. It reveals an increase of diameter at the input and a decrease (or focusing) at the output. The deformation is reversible and initial shape is completely recoverable on stopping the flow (See supplementary information S4).

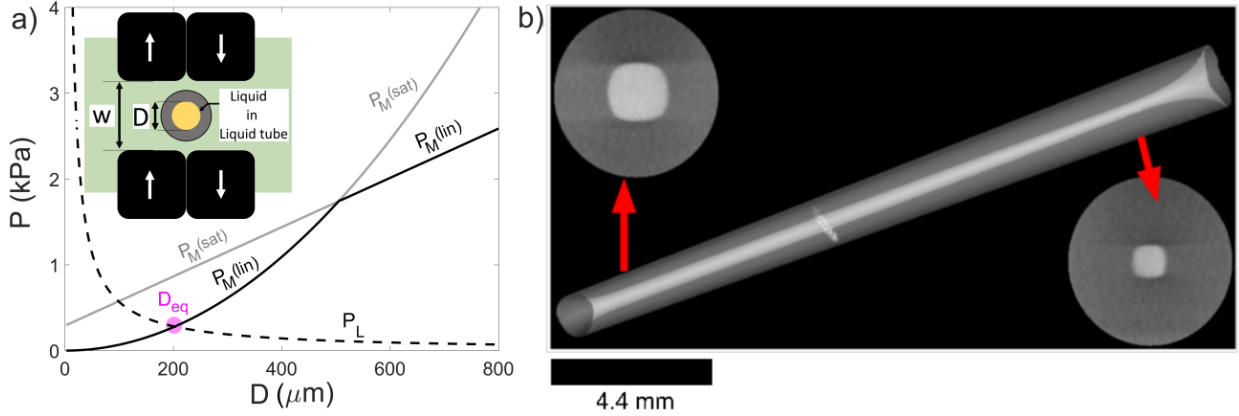

Figure 1. Deformable soft liquid antitube made of glycerol surrounded by APGE32 ferrofluid, viscosity ratio, $\eta_r = 0.65$. a) Magnetic design and the resulting pressures in the system. $P_M$ denotes the magnetic pressure (with $P_M(\text{lin})$ and $P_M(\text{sat})$ discriminating between the two limiting linear and saturation expressions of the bulk magnetic pressure, see details in the text) $P_L$ is the Laplace pressure. Inset shows the magnetic design. $D_{eq}$ is the static equilibrium diameter, $D$ and $W$ denote the diameter of the inside liquid tube and the spacing between magnet pair respectively. b) 3D X-ray tomography showing a deformed conical liquid tube at $Q = 75\ \mu\text{L/min}$. The cross sections emphasizes the difference between inlet (left) and outlet (right).

Reversible deformations of a glycerol antitube (central bright part Fig. 1b) were measured with three different ferrofluids encapsulation (APG1141, APGE32 and APG314 from Ferrotec). The main difference is the viscosity ratio ($\eta_r = \eta_a/\eta_f$) between glycerol and ferrofluids less than or greater than 1. For each $\eta_r$, three different no-flow diameters ($D_0$), with $D_0 > D_{eq}$, were set by adding different amounts of ferrofluid to the cavity first filled with glycerol. For each $D_0$, three different flow rates ($Q$) were applied using a syringe pump (Harvard apparatus PHD 2000). 3D tomography in Fig. 1b) and radiography (2D) imaging detailed in the supplementary information S3 and S4 confirms the stability and the reversibility of the observed conical shape of the antitube under flow. Fig 2a-c) shows the tracked antitube diameter measured as a function of $z$ (along the length of flow) by 2D X-ray imaging using edge detector algorithm [8] before flow (Fig. 2a) and during flow (Fig. 2b,c), with permanent magnets of length 40 mm at $z$ values between 5 and 45 mm. The curved shape near inlet and outlet relate to the effect of fringe magnetic field at the magnets ends. The deformation is more pronounced with increasing $Q$ and decreasing $D_0$, due to the increased fluid shear and the lower magnetic pressure when decreasing $D_0$. This hints towards a balance between viscous shear and magnetic pressure. If we consider the evolution of magnetic pressure, Laplace pressure and pressure driven

Stokes flow in the antitube, the axial gradient of the excess pressure responsible for the flow[4] is given by equation (1)

$$\frac{dP}{dz} = \frac{\partial}{\partial z}\left(\int_0^{H_{Max}} \mu_0 M(H) dH\right) + \frac{\partial}{\partial z}\left(\frac{1}{2}\mu_0 M_I^2\right) + \frac{\partial}{\partial z}(\sigma\kappa) = -128\eta_a Q/\pi D^4 \beta_D \qquad (1).$$

$\eta_a$ is the viscosity of antitube liquid (here glycerol), $Q$ the flow rate and $\beta_D$ the drag reduction factor due to the liquid-liquid interface[30].

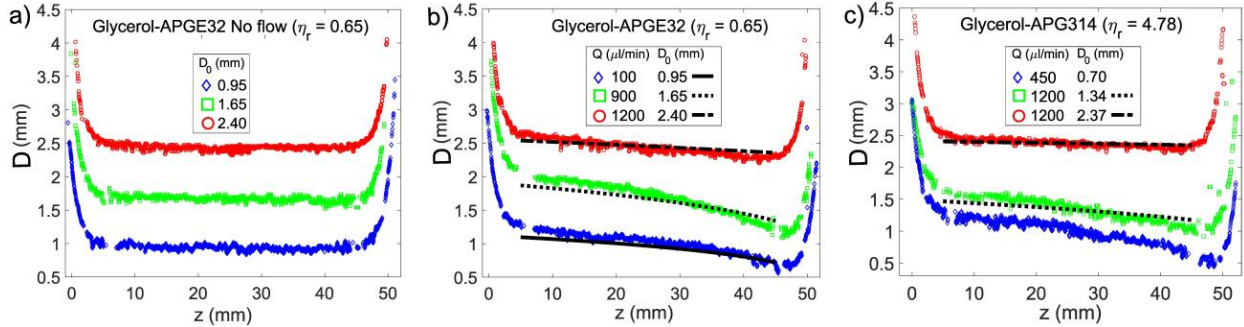

Figure 2. Deformation of liquid tube along its flow direction $z$. a) Diameter profile without flow for glycerol-APGE32 system, with three different $d_0$ (measured without considering inlet and outlet). b) Deformed diameter profile under flow rate $Q$. c) Deformed diameter profile for glycerol-APG314. $\eta_r$ is the viscosity ratio of the two liquids. Lines are analytical predictions from equation (2).

Since the deflected profile $D(z)$ has a small slope in the region excluding curved inlet and outlet, we neglect $D'''$ that scales as $D_0/L^3$, where $L$ is the length of the flow channel system (for full derivation, see supplementary information S5). This results into

$$D' = \frac{-32\eta_a Q}{\pi D^4 \beta\left[\frac{M_r}{\pi W}A[1+B] - \frac{\sigma}{D^2}\right]} \qquad (2)$$

where A and B are functions of $H$ at the interface (see explicit expressions in supplementary information S5). equation (2) is solved numerically for the conservation of ferrofluid volume, $\int_0^L D^2 dz = D_0^2 L$, see supplementary information S5. The resulting calculated interface profile $D(z)$ compares well with the experimental data in Fig. 2b) and Fig. 2c) for a rather wide range of $\eta_r$ values.

When the flow rate is increased beyond a critical value $Q_c$, an egress of ferrofluid occurs, as seen for the largest flow rate in Fig. 2c) where fitting with equation (2) becomes difficult.

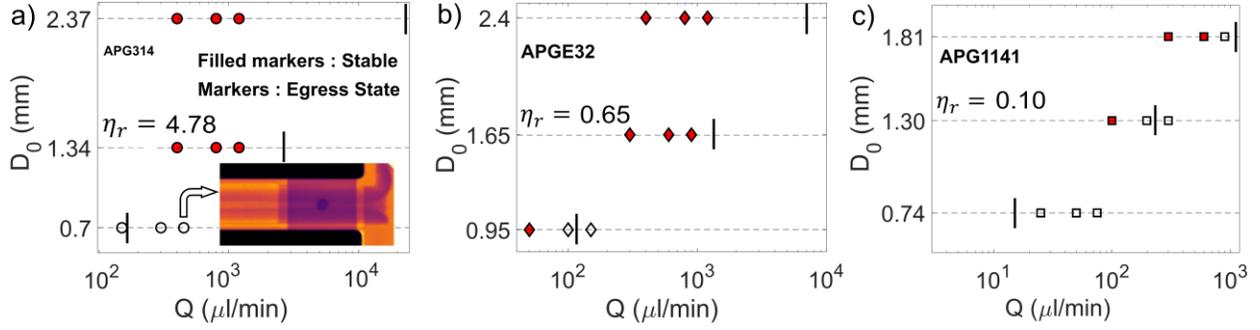

Figure 3. Stability range of antitubes flow, varying the diameter, the relative viscosity of the two liquids, and the flow rate. Filled markers and open markers denote experimentally stable and unstable egress ferrofluid states. Vertical line corresponds to $Q_c$ calculated analytically for RPI instability occurrence, using equation (5). a) Glycerol-APG314 with $\eta_r = 4.78$, inset shows a case where ferrofluid tries to escape. b) Glycerol-APGE32, $\eta_r = 0.65$, c) Glycerol-APG1141, $\eta_r = 0.10$

The inset of Fig. 3a), which corresponds to the image used for Fig. 2c) blue markers, illustrate the instability. A similar leak can also be observed for glycerol-APGE32, (see supplementary information S5). This indicates the occurrence of a cutoff flow rate above which the cylindrical structure becomes unstable, driven by the ferrofluid that escapes the magnetic force field. Experiments indicate qualitatively that $Q_c$ decreases with $D_0$ and $\eta_r$. This is expected as the Laplace pressure becomes more important at small diameters, and viscous shear increases when decreasing $\eta_r$, with the asymptotic case of $\eta_r \to 0$ corresponding to ferrofluid becoming solid-like.

Even though the weakest ferrofluid confining forces at the output are experimentally found to limit the stability of the system, we investigate the physics behind the limits of stability within the RPI framework. In the deformed state, stability requires that the viscous energy dissipation rate given by $\frac{dE}{dt} = \int_0^{d/2} 2\pi r \eta_a \left(\frac{r}{2\eta_1}\frac{\partial P}{\partial z}\right)^2 dr$, must be positive for all wave number ($k$) of perturbation.[31] Under perturbation of the type $R = R_{min} + \partial R_{min} \text{Cos} kz$ with $\partial R_{min}/R_{min} \ll 1$, with $R = \frac{D}{2}$, the radius of the antitube and $R_{min} = \frac{D_{min}}{2}$ the minimum radius of antitube (deformed) at the outlet (at z = L). We find that for linear magnetic media, $dE/dt$ obeys:

$$\frac{dE}{dt} = X\left[\frac{D_{min}^2 k^2}{4} - 2 + \frac{4D_{min}^3 \mu_0 \chi(\chi+1)M_r^2}{\sigma \pi^2 W^2}\right] \quad (3)$$

where $X$ is a positive multiplier (see details of the derivation in supplementary information S5). In the absence of magnetic forces, the last term in equation (3) will disappear and $\frac{dE}{dt}$ can be negative for small $k$. Hence the system becomes unstable as reported in the RPI literature[3,4].

However, in the presence of magnetic forces, $\frac{dE}{dt}$ is positive if:

$$\frac{4D_{min}^3 \mu_0 \chi(\chi+1)M_r^2}{\sigma \pi^2 W^2} - 2 > 0 \qquad (4)$$

resulting in a condition for stability

$$D_{min} > \frac{1}{2^{1/3}} \sqrt[3]{\frac{\sigma \pi^2 W^2}{\mu_0 \chi(\chi+1)M_r^2}} \qquad (5).$$

It can be rewritten as

$$D_{min} > 2(l_M^2/l_L) \qquad (6)$$

where, $l_M = \left(\eta_a Q w^2 L \pi / 2\mu_0 \chi(\chi+1) M_r^2 \beta_{D_0}\right)^{1/6}$ in the linear approximation the ferrofluid magnetization is understood as the deformation length scale that result from the balance between the magnetic and shear forces, and $l_L = \left(8\eta_a Q L/\sigma \pi \beta_{D_0}\right)^{1/3}$ is the length scale that expresses the ratio of viscous and surface tension (see supplementary information S5). Their simple ratio in equation (6) defines the stability condition, with $D_{min}$ is the diameter at the output, calculated from the deformation equation (equation (2)) and $\beta_{D_0}$ is the drag reduction parameter[30] calculated for $D_0$. Vertical lines in Fig. 3 are the calculated maximum permissible flow rate $Q_c$ obtained analytically using equation (2) and equation (5), such that $Q = Q_c$ for $D_{(z=L)} = D_{min}$. Note that most experiments show stability persisting very near the calculated flow limit, indicating that ferrofluid shearing effects were not exceedingly limiting our experiments stability range. One should emphasize the significant flow range where stability occurs, notably showing that stability is always kept at the smallest flow rates, in contrast to a standard RPI model.

One can take advantage of the magnetic forces stability against RPI to implement novel functional shapes with liquid-liquid interfaces. Fig. 4a) provides a clear illustration of the 'stabilization of the instability', where magnetic forces can simply be engineered by adding three spaced magnetic quadrupoles in series, capable to stabilize an oscillating cylindrical shape of liquid resembling the Rayleigh-Plateau instability (RPI) mode shape[3,4]. Fig. 4a) (right panel) shows the 3D tomography (See video V1 in supplementary information S6) of the spatially oscillating shape of liquid-in-liquid tube under flow. Such stable liquid cylindrical structures cannot be achieved by any other stabilization technique. It is reversibly stable at rest (Q = 0 µL/min), confirming that there is no lower flow limit for stability, as predicted by equation (5). This can open doors for complex reconfigurable microfluidics design. Fig. 4b) shows the conical liquid-in-liquid tube mimicking a nozzle shape with nozzle diameter of mm to $\mu m$ sizes (See video V2 for 3D tomography in supplementary information S6), with diameter reaching the microfluidic size range

($D_{min} = 96 \pm 6\ \mu m$) that can be focused (see supplementary information S6 for experimental setup), with no physics prohibiting smaller reachable diameters.

Beyond viscous dominated flow (Re < 1), the concept can be extended to create a high throughput, soft microfluidic circuit with Re >> 1 which does not shear away the ferrofluid with Reynolds numbers of several thousands (see supplementary information S6 for shearing details). Fig. 4c) shows the deformable microfluidics at micrometer sizes. To visualize the phenomena near instability, we used a 2D lubricated system, where ferrofluid is held only at two walls of a rectangular flow channel, thus making possible the transmission of light from top and we use optical microscopy to observe the dynamics of deformation and stability. Videos V3 and V4 in supplementary information S6 show stable and unstable microfluidic channels. One of the major limitation of flow focusing (liquid in liquid flow) is that it can only stabilize time invariant flow rate [3,32]. Magnetic confinements makes possible to overcome this limitation, as illustrated in Fig. 4d), where an oscillating water flow can be stabilized while time varying the pressure inlet (see video V5 and supplementary information S7 for details). Such unique capability could be applied to mimic unsteady flows encountered in living systems (see supplementary information S6).

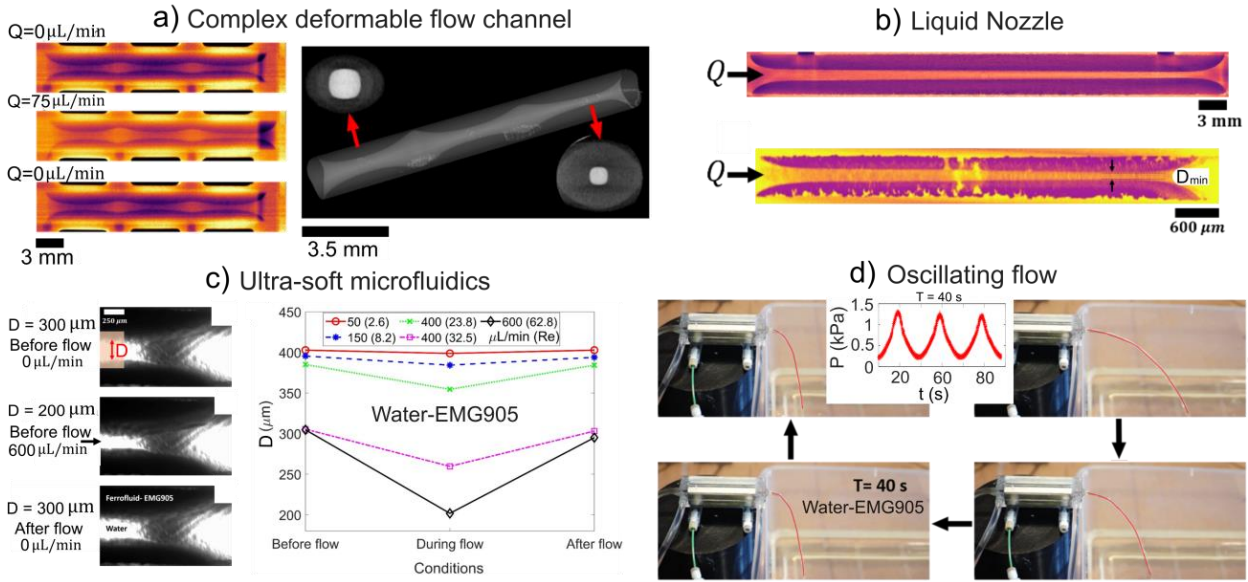

Figure 4. Towards deformable microfluidics of complex shapes. a) Stable oscillating (in space) shape liquid flow channel resembling RPI mode shape before flow (Q = 0 µL/min) , during flow (Q = 75 µL/min) and after flow (Q = 0 µL/min), 3D X-ray tomography (greyscale) of oscillating flow channel under flow, with cross sections at inlet and outlet. The permanent magnets are revealed by black areas. a) Liquid nozzle with Glycerol-APGE32, $D_{min}$ = 0.5 mm, $W$ = 6 mm (top) and Glycerol-EMG900 ($\eta_r$ = 10), $D_{min}$ = 96 $\pm$ 6 $\mu m$, $W$ = 0.9 mm. c) Soft 2D microfluidic channel with high Reynolds number flow using with water and ferrofluid EMG905 ($\eta_r$ = 0.2). d) Stable oscillating (in time) flow in a water antitube, snapshots at different time interval of the oscillating flow, under a 40 s period $T$ of the sinusoidal oscillation $P = P_0 + \Delta P\ \text{Sin}\left(\frac{2\pi t}{T}\right)$ of the pressure at the inlet , resulting into an oscillating flow ($Q_{max}$ at the peak of the curve is 250 ml/min). The antitube diameter is 2.4 mm (See supplementary information S6 for details).

In conclusion, the combination of magnetic pressure, viscous stress and Laplace pressure in a cylindrical liquid-in-liquid flow arrangement allows novel flow conditions, with the system stabilizing a liquid shape of cylindrical symmetry within a liquid environment. Our approach can possibly mitigate the failure mode of hydrodynamic flow focusing as well as completely avoid the dripping failure in co-flowing liquid streams [3] as stability holds down to static conditions. The balance of magnetic and viscous stress allows us to form a "*soft liquid-in-liquid nozzle*" of mm and sub-mm sizes which can find applications for example in bio-printing with minimum shear. We identify two length scales $l_M$ (ratio of viscous and magnetic parameters) and $l_L$ (ratio of viscous and surface tension parameter) such that $l_M$ governs the softness/deformations of the flow channel below a critical flow rate $Q_c$ which is defined by $l_M/l_L$. While retaining the benefits (continuous jet, low shear) and overcoming the limitations (continuous sheath flow, chemical manipulation of transported liquid) of other mentioned stability techniques, the balance of magnetic and fluid forces forms a new pathway to realize complex soft structures, stabilizing oscillating flows with liquid interface to be employed in a variety of applications and opening the door to unique bio-mimicking flow conditions.

Acknowledgements

We acknowledge numerous fruitful discussions with Dr. P. Dunne. We thank Prof. Gauthier, Antoine Egele, and Damien Favier of Institut Charles Sadron, Strasbourg, France for the use of the EasyTom X-ray Tomography facility. This project has received funding from the European Union's Horizon 2020 research and innovation programme under the Marie Skłodowska-Curie grant agreement MAMI No 766007 and QUSTEC No. 847471. We also acknowledge the support of the University of Strasbourg Institute for Advanced Studies (USIAS) and the Fondation Jean-Marie Lehn, as well as the support from IdEx Unistra (ANR 10 IDEX 0002), SFRI STRAT'US project (ANR 20 SFRI 0012) and EUR QMAT ANR-17-EURE-0024 under the framework of the French Investments for the Future Program. Support from the Institut Universitaire de France is also gratefully acknowledged (B.D. and T.H.)

References

1. Eggers, J. Nonlinear dynamics and breakup of free-surface flows. *Rev. Mod. Phys.* **69**, 865–930 (1997).

2. Rayleigh, Lord. On the Stability, or Instability, of certain Fluid Motions. *Proc. Lond. Math. Soc.* **s1-11**, 57–72 (1879).

3. Utada, A. S., Fernandez-Nieves, A., Stone, H. A. & Weitz, D. A. Dripping to Jetting Transitions in Coflowing Liquid Streams. *Phys. Rev. Lett.* **99**, 094502 (2007).

4. Haefner, S. *et al.* Influence of slip on the Plateau–Rayleigh instability on a fibre. *Nat. Commun.* **6**, 7409 (2015).

5. Vailati, A., Zinnato, L. & Cerbino, R. How Archer Fish Achieve a Powerful Impact: Hydrodynamic Instability of a Pulsed Jet in Toxotes jaculatrix. *PLOS ONE* **7**, e47867 (2012).

6. Jayaprakash, V., Costalonga, M., Dhulipala, S. & Varanasi, K. K. Enhancing the Injectability of High Concentration Drug Formulations Using Core Annular Flows. *Adv. Healthc. Mater.* **9**, 2001022 (2020).

7. Zhao, Z. *et al.* Breaking the symmetry to suppress the Plateau–Rayleigh instability and optimize hydropower utilization. *Nat. Commun.* **12**, 6899 (2021).

8. Blaeser, A. *et al.* Controlling Shear Stress in 3D Bioprinting is a Key Factor to Balance Printing Resolution and Stem Cell Integrity. *Adv. Healthc. Mater.* **5**, 326–333 (2016).

9. Hu, D. L., Chan, B. & Bush, J. W. M. The hydrodynamics of water strider locomotion. *Nature* **424**, 663–666 (2003).

10. Konrad, W. *et al.* The impact of raindrops on Salvinia molesta leaves: effects of trichomes and elasticity. *J. R. Soc. Interface* **18**, 20210676 (2021).

11. Ma, Y. *et al.* Relation between blood pressure and pulse wave velocity for human arteries. *Proc. Natl. Acad. Sci.* **115**, 11144–11149 (2018).

12. Skotheim, J. M. & Mahadevan, L. Soft Lubrication. *Phys. Rev. Lett.* **92**, 245509 (2004).

13. Bico, J., Roman, B., Moulin, L. & Boudaoud, A. Elastocapillary coalescence in wet hair. *Nature* **432**, 690–690 (2004).

14. Hosoi, A. E. & Mahadevan, L. Peeling, Healing, and Bursting in a Lubricated Elastic Sheet. *Phys. Rev. Lett.* **93**, 137802 (2004).

15. Arun Dev, A., Dey, R. & Mugele, F. Behaviour of flexible superhydrophobic striped surfaces during (electro-)wetting of a sessile drop. *Soft Matter* **15**, 9840–9848 (2019).

16. Barthlott, W. *et al.* The Salvinia Paradox: Superhydrophobic Surfaces with Hydrophilic Pins for Air Retention Under Water. *Adv. Mater.* **22**, 2325–2328 (2010).

17. Reis, P. M., Hure, J., Jung, S., Bush, J. W. M. & Clanet, C. Grabbing water. *Soft Matter* **6**, 5705–5708 (2010).

18. Feldmann, D., Das, R. & Pinchasik, B.-E. How Can Interfacial Phenomena in Nature Inspire Smaller Robots. *Adv. Mater. Interfaces* **8**, 2001300 (2021).

19. Leslie, D. C. *et al.* Frequency-specific flow control in microfluidic circuits with passive elastomeric features. *Nat. Phys.* **5**, 231–235 (2009).

20. Lam, E. W., Cooksey, G. A., Finlayson, B. A. & Folch, A. Microfluidic circuits with tunable flow resistances. *Appl. Phys. Lett.* **89**, 164105 (2006).

21. Box, F., Peng, G. G., Pihler-Puzović, D. & Juel, A. Flow-induced choking of a compliant Hele-Shaw cell. *Proc. Natl. Acad. Sci.* **117**, 30228–30233 (2020).

22. Unger, M. A., Chou, H.-P., Thorsen, T., Scherer, A. & Quake, S. R. Monolithic Microfabricated Valves and Pumps by Multilayer Soft Lithography. *Science* **288**, 113–116 (2000).

23. Thorsen, T., Maerkl, S. J. & Quake, S. R. Microfluidic Large-Scale Integration. *Science* **298**, 580–584 (2002).

24. Polygerinos, P. *et al.* Soft Robotics: Review of Fluid-Driven Intrinsically Soft Devices; Manufacturing, Sensing, Control, and Applications in Human-Robot Interaction. *Adv. Eng. Mater.* **19**, 1700016 (2017).

25. Pihler-Puzović, D., Illien, P., Heil, M. & Juel, A. Suppression of Complex Fingerlike Patterns at the Interface between Air and a Viscous Fluid by Elastic Membranes. *Phys. Rev. Lett.* **108**, 074502 (2012).

26. Verma, M. K. S. & Kumaran, V. A dynamical instability due to fluid–wall coupling lowers the transition Reynolds number in the flow through a flexible tube. *J. Fluid Mech.* **705**, 322–347 (2012).

27. Dunne, P. *et al.* Liquid flow and control without solid walls. *Nature* **581**, 58–62 (2020).

28. Coey, J. M. D., Aogaki, R., Byrne, F. & Stamenov, P. Magnetic stabilization and vorticity in submillimeter paramagnetic liquid tubes. *Proc. Natl. Acad. Sci.* **106**, 8811–8817 (2009).

29. Rosensweig, R. E. *Ferrohydrodynamics.* (Dover Publications, Incorporated, 2013).

30. Dev, A. A., Dunne, P., Hermans, T. M. & Doudin, B. Fluid Drag Reduction by Magnetic Confinement. *Langmuir* **38**, 719–726 (2022).

31. Chandrashekhar, S. *Hydrodynamic and hydromagnetic stability*. (Dover Publication).

32. Gañán-Calvo, A. M., González-Prieto, R., Riesco-Chueca, P., Herrada, M. A. & Flores-Mosquera, M. Focusing capillary jets close to the continuum limit. *Nat. Phys.* **3**, 737–742 (2007).